\documentclass[letterpaper,twocolumn,pra,aps,showpacs,superscriptaddress,
floatfix ]{revtex4-1}
\usepackage[latin1]{inputenc}
\usepackage{bm}
\usepackage[usenames]{color}
\usepackage{multirow}
\usepackage{amssymb}
\usepackage{amsbsy}
\usepackage{amsmath}
\usepackage{stmaryrd}
\usepackage{graphicx}
\usepackage{epsfig}
\usepackage{placeins}
\usepackage{ulem}
\usepackage{bbold}
\usepackage{braket}
\usepackage{blindtext}
\usepackage[colorlinks,linkcolor=blue,citecolor=blue,urlcolor=blue]{hyperref}

\makeatletter

\begin{document}

\title{Exponential damping induced by random and realistic perturbations}

\author{Jonas Richter}
\email{jonasrichter@uos.de}
\affiliation{Department of Physics, University of Osnabr\"uck, D-49069 
Osnabr\"uck, Germany}

\author{Fengping Jin}
\affiliation{Institute for Advanced Simulation, J\"ulich Supercomputing Centre,
Forschungszentrum J\"ulich, D-52425 J\"ulich, Germany}

\author{Lars Knipschild}
\affiliation{Department of Physics, University of Osnabr\"uck, D-49069 
Osnabr\"uck, Germany}

\author{Hans De Raedt}
\affiliation{Zernike Institute for Advanced Materials, University of Groningen,
NL-9747AG Groningen, The Netherlands}

\author{Kristel Michielsen}
\affiliation{Institute for Advanced Simulation, J\"ulich Supercomputing Centre,
Forschungszentrum J\"ulich, D-52425 J\"ulich, Germany}
\affiliation{RWTH Aachen University, D-52056 Aachen, Germany}

\author{Jochen Gemmer}
\email{jgemmer@uos.de}
\affiliation{Department of Physics, University of Osnabr\"uck, D-49069 
Osnabr\"uck,
Germany}

\author{Robin Steinigeweg}
\email{rsteinig@uos.de}
\affiliation{Department of Physics, University of Osnabr\"uck, D-49069 
Osnabr\"uck, Germany}

\date{\today}


\begin{abstract}

Given a quantum many-body system and the expectation-value dynamics of some 
operator, we study how this reference dynamics is altered due to a 
perturbation of the system's Hamiltonian. Based on projection operator 
techniques, we unveil that if the perturbation exhibits a random-matrix 
structure in the eigenbasis of the unperturbed Hamiltonian, then this 
perturbation effectively leads to an exponential damping of the original 
dynamics.
Employing a combination of dynamical quantum typicality and numerical 
linked cluster expansions, we demonstrate that our 
theoretical findings for random matrices can, in some cases, be relevant for 
the dynamics of realistic quantum 
many-body models as well. Specifically, we study the decay of current 
autocorrelation functions in spin-$1/2$ ladder systems, where the rungs of the 
ladder are treated as a perturbation to the otherwise uncoupled legs. We find a 
convincing agreement between the exact dynamics and the lowest-order prediction 
over a wide range of interchain couplings. 

\end{abstract}

\maketitle


\section{Introduction}\label{Sec::Intro}

Understanding the dynamics of interacting quantum many-body systems is 
notoriously challenging. While their complexity grows exponentially in the 
number of degrees of freedom, the strong correlations between the 
constituents often prohibit any exact solution. Although much progress has been 
made due to the development of powerful numerical machinery 
\cite{Schollwoeck2011} and the advance of controlled experimental platforms 
\cite{Bloch2008, Langen2015}, the detection of general 
(i.e.\ universal) principles which underlie the emerging many-body dynamics is 
of fundamental importance \cite{Polkovnikov2011}. To this end, a remarkably 
successful 
strategy in the past has been the usage of random-matrix ensembles instead of 
treating the full many-body problem. Ranging back to the 
description of 
nuclei spectra \cite{Brody1981} and of quantum chaos in systems with classical 
counterparts \cite{Bohigas1984}, random-matrix theory also forms the backbone 
of the celebrated eigenstate thermalization hypothesis (ETH) \cite{Deutsch1991, 
Srednicki1994, Rigol2005}, which provides 
a microscopic explanation for the emergence of thermalization in 
isolated quantum systems. More recently,
random-circuit models have led to new insights into the scrambling of 
information and the onset of hydrodynamic transport in quantum systems 
undergoing unitary time evolution \cite{Keyserlingk2018, Nahum2018, 
Khemani2018}. 

Concerning the out-of-equilibrium dynamics of quantum many-body systems, a 
particularly intriguing question is how the expectation-value dynamics of some 
operator is altered if the system's Hamiltonian is modified by a (small or 
strong) 
perturbation. Clearly, the effect of such a perturbation in a 
(integrable or chaotic) system can be manifold. In the context of 
prethermalization 
\cite{Berges2004, Moeckel2008, Bertini2015, Mori2018, Reimann2019, 
Mallayya2019}, the perturbation breaks a conservation law of 
the (usually 
integrable) reference Hamiltonian, leading to a separation of time scales, where 
the system stays close to some long-lived nonthermal state, before eventually 
giving in to its thermal fate at much longer times. Moreover, in the 
study of echo protocols, time-local perturbations have been shown to entail 
irreversible quantum dynamics \cite{Schmitt2018}, analogous to the 
butterfly effect in classical chaotic systems. Furthermore, the 
observation that some types of temporal relaxation, such as the exponential 
decay, are more common than others can be traced back to their enhanced 
stability against perturbations \cite{Knipschild2018}.  

In this paper, we consider a closed quantum many-body 
system ${\cal H}_0$ which is affected by a perturbation ${\cal V}$, such that 
the total Hamiltonian takes on the form 
\begin{equation}\label{Eq::Decomp}
{\cal H} = {\cal H}_0 + \lambda {\cal 
V}\ , 
\end{equation}
where $\lambda$ denotes the strength of the perturbation.
Given the 
expectation-value dynamics of some operator ${\cal O}$ in the unperturbed 
system, 
\begin{equation}
\langle {\cal O}(t) \rangle_{{\cal H}_0} = \text{Tr}[{\cal 
O}\rho(t)]\ ,  
\end{equation}
where $\rho(t) = 
e^{-i{\cal H}_0t}\rho(0)e^{i{\cal H}_0t}$ [and $\rho(0)$ is 
a mixed or pure out-of-equilibrium initial state], we explore the question how 
$\langle {\cal O}(t) \rangle_{{\cal H}_0}$ is altered due to the presence of 
the perturbation, i.e., 
if $\rho(0)$ now evolves with respect to the full Hamiltonian ${\cal H}$.
Employing the time-convolutionless (TCL) projection operator method 
\cite{Chaturvedi1979, Breuer2007}, we unveil 
for the idealized case of
${\cal V}$ exhibiting a random-matrix
structure in the eigenbasis of ${\cal H}_0$, that 
such a perturbation effectively leads to
an exponential damping of the original dynamics (see also \cite{Dabelow2019, 
Nation2019}),  
\begin{equation}\label{Eq::ExpDec}
 \langle {\cal O}(t) \rangle = \langle {\cal O}(t) \rangle_{{\cal H}_0} 
e^{-\lambda^2 \gamma t}\ ,
\end{equation}
where the damping rate $\gamma$ depends on the microscopic properties of ${\cal 
H}_0$ and ${\cal V}$. 
In order to illustrate that our 
analytical findings for random matrices can indeed be relevant for the 
dynamics of realistic 
quantum many-body systems, we numerically study the decay of current 
autocorrelation functions in spin-1/2 ladder models, where the rungs of the 
ladder are
treated as a perturbation to the otherwise uncoupled legs. Especially for 
small to intermediate values of the interchain coupling, we find a convincing 
agreement between the exact dynamics and the leading-order prediction.

This paper is structured as follows. 
In Sec.\ \ref{Sec::TCL}, we discuss the TCL formalism and present an 
analytical derivation of Eq.\ \eqref{Eq::ExpDec} for the case of ${\cal V}$ 
having an ideal random-matrix structure in the eigenbasis of ${\cal H}_0$. 
Next, in Sec.\ \ref{Sec::NumIllus}, we test the applicability of 
Eq.\ \eqref{Eq::ExpDec} by studying the real-time dynamics for more realistic 
models and perturbations using an efficient combination of dynamical quantum 
typicality and numerical linked cluster expansions. We summarize and conclude 
in Sec.\ \ref{Sec::Conclu}. 


\section{Projection operator approach to ideal random-matrix 
models}\label{Sec::TCL}

\subsection{Derivation of the main result}

Let us now derive Eq.\ \eqref{Eq::ExpDec} within the framework of the 
TCL projection operator technique. 
In the TCL formalism, 
the decomposition of the Hamiltonian \eqref{Eq::Decomp} is usually done in such 
a way that the observable 
of interest (here the operator ${\cal O}$) is either conserved in the 
unperturbed system or only shows slow 
dynamics under evolution with ${\cal H}_0$.
Moreover, in order to apply the TCL formalism, one needs to define a suitable  
projection operator ${\cal P}^2 = {\cal P}$, which projects onto the relevant 
degrees of freedom. For a comprehensive review, see, e.g., 
\cite{Chaturvedi1979, Breuer2007}. 

In order to simplify the upcoming derivation, let us assume that ${\cal 
H}_0$ 
has a very high and almost uniform density of states 
\begin{equation}
 \Omega \approx 1/\Delta \omega\ , 
\end{equation} 
where $\Delta \omega$ is the mean level spacing. Moreover, we 
shift the true eigenvalues of ${\cal H}_0$ slightly, such that they result as 
$E_\omega = \Delta \omega \cdot \omega$, with $\omega$ being an integer. 
Although these conditions are not necessarily fulfilled 
for a given system, we expect corrections to our results to be irrelevant on 
time scales $t\ll 
2\pi/\Delta\omega$.

To begin with, we define a Fourier component of the operator ${\cal O}$ 
in the 
eigenbasis of ${\cal H}_0$,
\begin{equation}
{\cal O}_{\omega} = \frac{1}{\sqrt{z_\omega}}\sum_\eta \ket{\eta} {\cal 
O}_{\eta,\eta+\omega} \bra{\eta + \omega} + \text{h.c.}\ ,
\end{equation}
with normalization $z_\omega = 2\sum_\eta |{\cal O}_{\eta,\eta+\omega}|^2$,
and construct a set of corresponding projection operators ${\cal P}_\omega$, 
which project onto the relevant part of the density matrix $\rho(t)$, 
\begin{equation}
 {\cal 
P}_\omega \rho(t) = \text{Tr}[\rho(t){\cal O}_\omega]{\cal O}_\omega\ .
\end{equation}
Note that, from the definition of ${\cal O}_\omega$, 
it immediately follows that
\begin{equation}
 \langle {\cal O}(t) \rangle = \sum_\omega \sqrt{z_\omega} \langle {\cal 
O}_\omega (t) \rangle\ ,
\end{equation}
 which holds for any ${\cal H}$, and in the special 
case $\lambda =0$, we can further write 
\begin{equation}
 \langle {\cal O}(t) 
\rangle_{{\cal H}_0} = \sum_\omega \sqrt{z_\omega} \langle {\cal O}_\omega(0) 
\rangle \cos(\Delta \omega\cdot \omega t)\ .
\end{equation}
 In particular, the  
dynamics of ${\cal O}_\omega$ in the Schr\"odinger picture and in the 
interaction picture (subscript 
$I$) are related by 
\begin{equation}
 \text{Tr}[{\cal O}_{\omega} \rho(t)] = \text{Tr}[{\cal 
O}_{\omega} \rho_I(t)]\cos(\Delta \omega \cdot \omega t)\ . 
\end{equation}

Next, we focus on initial states $\rho(0)$,  which fulfill ${\cal P}_\omega 
\rho(0) = \rho(0)$ \cite{Breuer2007}. 
For such initial states, the TCL framework then yields a 
time-local equation for 
${\cal O}_{I,\omega}(t) = \text{Tr}[{\cal O}_{\omega} \rho_I(t)]$, 
comprising a systematic perturbation expansion in powers of $\lambda$ 
\cite{Breuer2007},
\begin{equation}\label{Eq::TCL}
\dot{{\cal O}}_{I,\omega}(t) = -
\gamma_\omega(t) 
{\cal O}_{I,\omega}(t)\ ;\ \ \gamma_\omega(t) = \sum_n 
\lambda^n 
\gamma_{\omega,n}(t)\ ,  
\end{equation}
where the $\gamma_{\omega,n}(t)$ are time-dependent rates of $n$-th order.
Due to our choice of the ${\cal P}_\omega$, all odd orders of 
this expansion vanish (as it is often the case in the TCL framework 
\cite{Breuer2007}), and the leading-order 
term is 
\begin{align}
 \gamma_{\omega,2}(t) &= \int_0^t \text{d}t' K_{\omega,2}(t')\ , 
\end{align}
with 
\begin{align}
K_{\omega,2}(t) = \text{Tr}\left\lbrace i[{\cal O}_\omega,{\cal V}_I(t)]i[ 
{\cal O}_\omega,{\cal V}]\right\rbrace\ \label{Eq::Gamma2},
\end{align}
and ${\cal V}_I(t) = e^{i{\cal H}_0t} {\cal V} e^{-i{\cal H}_0t}$. Here, the 
second-order kernel can be rewritten as $K_{\omega,2}(t) = {\widetilde 
K}_{\omega,2}(t) + {\widehat K}_{\omega,2}(t)$, with  
\begin{align}
 {\widetilde K}_{\omega,2}(t) &= \text{Tr}\left[-{\cal O}_\omega {\cal V}_I(t) 
{\cal 
O}_\omega {\cal V} - {\cal V}_I(t) {\cal O}_\omega {\cal V} {\cal 
O}_\omega \right] \, , \label{Eq::K1}\\
 {\widehat K}_{\omega,2}(t) &= \text{Tr}\left[ {\cal O}_\omega {\cal V}_I(t) 
{\cal V} {\cal O}_\omega + {\cal 
O}_\omega {\cal V} {\cal V}_I(t) {\cal O}_\omega \right]\ . \label{Eq::K2}
\end{align}
Let us stress that we made no assumptions on the specific 
form of the perturbation ${\cal V}$ up to this point. 

For the {\it idealized} case of ${\cal V}$ being an 
entirely random (and possibly banded) matrix in the eigenbasis of 
the unperturbed system ${\cal H}_0$, it is possible to derive an 
analytic expression for the leading-order rate $\gamma_2(t)$. Focusing on this 
case, 
the terms in \eqref{Eq::K1} consist of sums in which each addend carries a 
product of two uncorrelated random numbers. If the random numbers have mean 
zero, these sums should be negligible, 
\begin{equation}
 {\widetilde K}_{\omega,2}(t) \approx 
0\ . 
\end{equation}
 In contrast, the terms in 
\eqref{Eq::K2} do contribute, and we find
\begin{align}
 {\widehat K}_{\omega,2}(t) &= \frac{4}{z_\omega} \sum_{\eta, \kappa} |{\cal 
V}_{\kappa,\eta}|^2 
|{\cal O}_{\eta,\eta + \omega}|^2 \cos\left[(\kappa - \eta)\Delta \omega 
t\right] \nonumber \\
&\approx \frac{4\Omega \overline{v^2}}{z_\omega} \sum_\eta 
\int_{-W}^{W}|{\cal O}_{\eta,\eta+\omega}|^2 \cos(\chi t) \, \text{d}\chi 
\label{Eq::1} \\
&=2\Omega \overline{v^2} \int_{-W}^{W} \cos(\chi t) \, \text{d}\chi = 
4\Omega  \overline{v^2} \, \frac{\sin(Wt)}{t}\ \label{Eq::2}.
\end{align}
Several comments are in order. Since ${\cal V}$ is a random matrix, we have 
approximated in Eq.\ \eqref{Eq::1} all squared individual matrix elements by 
their averages $\overline{v^2}$, i.e., $|{\cal V}_{\kappa,\eta}|^2 
\approx \overline{v^2}$. Furthermore, we have used an index shift $\kappa \to 
\chi + \eta$ and converted the original sum over $\kappa$ to an integral, where 
$W$ denotes the half-bandwidth of ${\cal V}$. From \eqref{Eq::1} to 
\eqref{Eq::2}, we have exploited that sum and integral can be evaluated 
independently and used the definition of $z_\omega$.  
Inserting \eqref{Eq::2} into
the definition of $\gamma_{\omega,2}(t)$
yields 
\begin{equation}
 \gamma_{\omega,2}(t) \approx 4\Omega\overline{v^2}\int_0^t 
\text{d}t'
\frac{\sin(W t')}{t'} \approx 2\pi\Omega\overline{v^2}\ ,
\end{equation}
for times $t \gg 
\pi/W$, and we abbreviate 
\begin{equation}
 \gamma = 
2\pi\Omega\overline{v^2}\ . 
\end{equation}
From the inspection of Eq.\ 
\eqref{Eq::TCL} it then follows that   
${\cal 
O}_{I,\omega}(t) = 
\langle {\cal O}_\omega(0) \rangle e^{-\lambda^2 \gamma t}$, and a 
transformation back to the Schr\"odinger picture  
yields our main result in Eq.\ \eqref{Eq::ExpDec}. Note that $\gamma$ is of the 
very same form as a rate describing the transition out of an initially fully 
populated eigenstate $|\eta\rangle$ of ${\cal H}_0$, as induced by ${\cal 
V}$, calculated from Fermi's Golden Rule. 

\subsection{Discussion of the main result}

Let us discuss our main result \eqref{Eq::ExpDec} in some more detail. 
First, we note that Eq.\ \eqref{Eq::ExpDec} is consistent with very recent 
findings in Refs.\ \cite{Dabelow2019, 
Nation2019}, although the employed approaches to arrive at this result have 
been very different. While the approaches in Refs.\ \cite{Dabelow2019, 
Nation2019} rely heavily on the concept of the perturbations being effectively 
represented by random matrices, their results on dynamics technically address 
averages over ensembles of random matrices. However, either by relying on 
``self-averaging'' \cite{Nation2019} or as the result of a detailed calculation 
\cite{Dabelow2019}, the outcome for a specific perturbation is expected to be 
very close to the ensemble average. On the contrary, since the present analysis 
is based on projection operator techniques, it is in principle applicable to 
any specific (matrix)form of the perturbation. The result of this technique is 
given as a perturbation series containing all orders of the interaction 
strength $\lambda$, cf.\ Eq.\ \eqref{Eq::TCL}. However, even in
leading order, the evaluation is in general rather involved. 

Next, we note that our derivations within the TCL approach are rigorous for an 
idealized random perturbation ${\cal V}$ up 
to second order in the 
perturbation strength. This truncation relies on ${\cal V}^2$ being dominantly 
diagonal (in the eigenbasis of ${\cal H}_0$). Random matrices also, but not 
exclusively, exhibit this feature \cite{Bartsch2008}.

While we have derived Eq.\ \eqref{Eq::ExpDec} for an idealized model and 
perturbation, this does not 
necessarily
exclude the possibility that this equation is relevant 
also beyond such idealized cases. 
For instance, the ETH assumes a (almost) random-matrix structure of physical 
operators in the 
eigenbasis of generic Hamiltonians \cite{Srednicki1994, Rigol2005}, 
as numerically verified for various models 
\cite{Beugeling2015, Mondaini2017, Richter2019_3}. 
In fact, in the upcoming Sec.\ \ref{Sec::NumIllus}, we numerically 
illustrate that Eq.\ \eqref{Eq::ExpDec} is indeed 
also applicable to understand 
the 
dynamics of certain realistic quantum many-body systems and perturbations.
In this context, let us add that the phenomenon of an exponential damping has 
been found for an even wider range of realistic models, for instance in 
Refs.\ \cite{Flesch2008, Trotzky2012, Barmettler2009, Balzer2015}.

Nevertheless, we should stress that for a given model and perturbation, it is 
\textit{a priori} certainly questionable whether a unitary basis transformation 
of the (originally nonrandom) perturbation can indeed yield entirely 
uncorrelated matrix elements. 
One criterion to check whether or not our arguments for random matrices 
also hold for realistic models is the evaluation of higher-order 
corrections.
For example, the fourth-order rate $\gamma_4(t)$ in the TCL formalism reads
\begin{eqnarray}
\gamma_{4}(t) &=& \int_{0}^{t} \! \text{d}t_1 \int_{0}^{t_1} \!
\text{d}t_2 \int_{0}^{t_2} \text{d}t_3 \label{r4} \\[0.1cm]
&& \! K_2(t-t_1) \; K_2(t_2-t_3) \nonumber \\[0.1cm]
&+& \! K_2(t-t_2) \; K_2(t_1-t_3) \nonumber \\[0.1cm]
&+& \! K_2(t-t_3) \; K_2(t_1-t_2) \nonumber \\[0.1cm]
&-& \! \text{Tr}\lbrace [ [{\cal O}, {\cal V}_{I}(t_1)],
{\cal V}_{I}(t) ] \, [ [{\cal O}, {\cal V}_{I}(t_3)],
{\cal V}_{I}(t_2) ] \rbrace\, \nonumber , 
\end{eqnarray}
where we have dropped the subscript $\omega$ for simplicity. If one finds that 
$\gamma_4(t)$ 
is significantly smaller 
than $\gamma_2(t)$ on the time scale of relaxation, this could be interpreted 
as an indication that ${\cal V}$ 
(in the eigenbasis of ${\cal 
H}_0$) is sufficiently well describable by a (pseudo)random matrix. Note that, 
in practice, the (numerical or analytical) evaluation of Eq.\ \eqref{r4} is 
considerably more 
difficult compared to the second-order rate in Eq.\ \eqref{Eq::Gamma2} 
\cite{Steinigeweg2013}. 

While 
the detailed analysis of correlations between matrix elements is beyond the 
scope of the present paper, let us note that one can find various 
realistic models where $\gamma_4(t) \ll \gamma_2(t)$ \cite{Steinigeweg2011},  
while there are naturally also other models where this 
property as well as Eq.\ \eqref{Eq::ExpDec} do not hold anymore.  
One simple example would be 
the case where observable and perturbation commute, i.e., $[{\cal O},{\cal V}] 
= 0$, 
and the second-order rate $\gamma_2(t)$ vanishes exactly.
Another example where our framework necessarily fails by construction would be 
given by reversing the roles of ${\cal H}_0$ and ${\cal H}$, i.e., by defining 
a new 
unperturbed Hamiltonian as ${\cal H}_0^\prime = {\cal H}_0 + \lambda {\cal V}$, 
which is then perturbed by ${\cal V}' = -\lambda{\cal V}$ such that ${\cal H}' 
= {\cal 
H}_0$.   


\section{Numerical Illustration for quantum many-body 
systems}\label{Sec::NumIllus}

Let us now numerically illustrate that our main result \eqref{Eq::ExpDec} can 
be relevant for the dynamics of realistic quantum many-body systems. First, in 
Sec.\ \ref{Sec::ModelObs}, we introduce the specific model and observable under 
consideration. In Sec.\ \ref{Sec::Numerics}, we then discuss our numerical 
approach which is used to study the real-time dynamics of the unperturbed and 
the perturbed system in Secs.\ \ref{Sec::Unperturbed} and \ref{Sec::Perturbed}, 
respectively. Moreover, we comment 
on the matrix 
structure of the realistic perturbation in the eigenbasis of ${\cal H}_0$ in 
Sec.\ \ref{Sec::MatrixStructure}. 

\subsection{Model and Observable}\label{Sec::ModelObs}

We study a (quasi-)one-dimensional spin-$1/2$ lattice 
model with 
ladder geometry \cite{Zotos2004, Jung2006, Znidaric2013,
Karrasch2015, Steinigeweg2016}, where the rung part of the ladder is treated
as a perturbation to the otherwise uncoupled legs, i.e., the Hamiltonian reads
${\cal H} = J_\parallel {\cal H}_0 + J_\perp {\cal V}$, with
\begin{equation}\label{Eq::Ham}
 {\cal H}_0 =  \sum_{l=1}^L \sum_{k=1}^2 
 S_{l,k}^x S_{l+1,k}^x + S_{l,k}^y S_{l+1,k}^y + \Delta S_{l,k}^z S_{l+1,k}^z
\end{equation}
and
\begin{equation}\label{Eq::Perturb}
 \quad {\cal V} = \sum_{l = 1}^L 
 S_{l,1}^x S_{l,2}^x + S_{l,1}^y S_{l,2}^y + \Delta S_{l,1}^z S_{l,2}^z\ . 
\end{equation}
Here, ${\bf S}_{l,k} = (S_{l,k}^x, S_{l,k}^y, S_{l,k}^z)$ are spin-$1/2$ 
operators, $J_\parallel$ ($J_\perp)$ is the coupling constant on the legs 
(rungs), and $L$ denotes the length of the ladder. Moreover, the anisotropy 
$\Delta$ is chosen to be either $\Delta = 0$ (XX ladder) or $\Delta = 1$ 
(XXX ladder). 
While, for $J_\perp = 
0$, ${\cal H}$ consists of two separate chains and is integrable, this 
integrability is broken for any $J_\perp \neq 0$. 

For this model, let us study the 
current autocorrelation function
\begin{equation}
 C(t) = \frac{\langle j(t) j \rangle_\text{eq}}{L} = 
  \frac{\text{Tr}[\rho_\text{eq} j(t)j]}{L}\ , 
\end{equation}
where $\rho_\text{eq} = e^{-\beta {\cal H}}/{\cal Z}$ is the canonical density 
matrix, $\beta = 1/T$ denotes the inverse temperature, and $j(t) = e^{i{\cal 
H}t} j e^{-i{\cal H}t}$. Moreover, the spin-current operator 
$j$ follows from a lattice continuity equation \cite{Heidrichmeisner2007}, and 
is given by
\begin{equation}
 j = J_\parallel \sum_{l=1}^L \sum_{k=1}^2 (S_{l,k}^x S_{l+1,k}^y - 
 S_{l,k}^y S_{l+1,k}^x )\ .
\end{equation}
(Note that $j$ is independent of the perturbation ${\cal V}$.)
Specifically, we here focus on the case
of infinite temperature $\beta = 0$ ($\rho_\text{eq} = \mathbb{1}/4^L$), and 
the correlation function $C(t)$ can be interpreted as the expectation-value 
dynamics 
\begin{equation}\label{Eq::CtExpv}
 C(t) = \langle j(t) \rangle = \text{Tr}[j\rho(t)]\ , 
\end{equation}
resulting from an initial 
state 
\begin{equation}
 \rho(0) \propto \mathbb{1} + \varepsilon j\ ,\ \quad \text{Tr}[\rho(0)] = 1\ , 
\end{equation}
with $\varepsilon$ sufficiently small. Given the decomposition in Eqs.\ 
\eqref{Eq::Ham} and \eqref{Eq::Perturb} and  choosing a simple projection onto 
the current $j$,
\begin{equation}\label{Projecj}
 {\cal P}\rho(t) = \frac{1}{4^L} + \frac{\text{Tr}[j\rho(t)]}{\text{Tr}[j^2]}j\ 
, 
\end{equation}
the TCL formalism can be readily applied to Eq.\ 
\eqref{Eq::CtExpv}. As a consequence, the derivations 
outlined in Sec.\ \ref{Sec::TCL} for the expectation-value dynamics 
$\langle {\cal O}(t)\rangle$ carry over to the high-temperature correlation 
function $C(t)$, which allows us to test whether or not our results for random 
matrices are relevant for this more realistic setting.

\subsection{Numerical approach}\label{Sec::Numerics}

The correlation function $C(t)$ is an important quantity 
in the context of transport. Despite the integrability of ${\cal H}_0$, the 
dynamics of $C(t)$ is nontrivial
even for $J_\perp = 0$ \cite{Bertini2020}.
While $C(t)$ has been numerically studied by 
various methods \cite{Sirker2009, Karrasch2012, Karrasch2015, Richter2019_2}, 
we here rely 
on a 
combination of dynamical 
quantum typicality (DQT) \cite{Gemmer2004, Popescu2006, Goldstein2006, 
Reimann2007, Bartsch2009, Hams2000, Iitaka2003, Sugiura2013, Elsayed2013, 
Steinigeweg2014, Steinigeweg2015} and numerical linked cluster expansions 
(NLCE) \cite{Tang2013, Mallayya2018}, recently put forward by two of us 
\cite{Richter2019}. 

\subsubsection{Dynamical quantum typicality}\label{Sec::DQT}

On the one hand, the concept of DQT relies on the fact that a single pure 
quantum state can imitate the full statistical ensemble.
Specifically, for $\beta = 0$,  $C(t)$ can be written as a scalar product with 
the two 
pure states \cite{Elsayed2013, Steinigeweg2014}
\begin{align}
 \ket{\psi(t)} &= e^{-i{\cal H}t}j\ket{\varphi}\ , \\
\ket{\varphi(t)} &= e^{-i{\cal H}t} \ket{\varphi}\ , 
\end{align}
according to
\begin{equation}
 C(t) = \frac{\bra{\varphi(t)} j 
\ket{\psi(t)}}{L\braket{\varphi|\varphi}} + \epsilon\ ,  
\end{equation}
where the reference pure state $\ket{\varphi}$ is randomly drawn (Haar measure 
\cite{Bartsch2009})
from the full Hilbert space with dimension $D = 4^L$.
Importantly, the statistical error $\epsilon = \epsilon(\ket{\varphi})$
vanishes as $\epsilon \propto 1/\sqrt{D}$ (for 
$\beta = 0$), and the approximation 
becomes very accurate already for moderate values of $L$.
Since the time evolution of pure states can be 
conveniently evaluated by iteratively solving the  Schr\"odinger 
equation, e.g., by means of
fourth-order Runge-Kutta \cite{Elsayed2013, Steinigeweg2014} or Trotter 
decompositions \cite{deReadt2006},
it is possible to treat large
$D$, out of reach for standard exact diagonalization (ED).

\subsubsection{Numerical linked cluster expansion}

On the other hand, NLCE provides a means to obtain $C(t)$ directly in the 
thermodynamic limit $L \to \infty$. Specifically, the current autocorrelation 
is calculated as the sum of contributions from all connected clusters which can 
be embedded on the lattice \cite{Tang2013},
\begin{equation}\label{Eq::NLCE}
 \frac{\langle j(t) j \rangle_\text{eq}}{L} 
= \sum_{c} {\cal L}_c W_c(t)\ , 
\end{equation}
where $W_c(t)$ is the weight of cluster $c$ with multiplicity ${\cal L}_c$. The 
notion of a cluster here refers to a finite number of lattice sites which are 
coupled by the respective Hamiltonian.
In fact, for a one-dimensional chain geometry [and also a 
(quasi-)one-dimensional ladder system] the identification 
of clusters becomes very simple. Namely, clusters are just 
chains (or ladders) of finite length. More 
details can be found in 
Refs.\ \cite{Tang2013, Mallayya2018, Richter2019}. Moreover, since there is 
only one distinct cluster for a given cluster size, we have ${\cal L}_c = 1$ 
in Eq.\ \eqref{Eq::NLCE}. 

Given a cluster $c$, its weight $W_c(t)$ is obtained by the so-called 
inclusion-exclusion principle,
\begin{equation}\label{Eq::InkluExklu}
 W_c(t) = \langle j(t) j \rangle_\text{eq}^{(c)} - \sum_{s \subset c} W_s(t)\ , 
\end{equation}
where $\langle j(t) j \rangle_\text{eq}^{(c)}$ denotes the (extensive) current 
autocorrelation evaluated on the cluster $c$ (with open boundary conditions). 
Furthermore, the sum in Eq.\ \eqref{Eq::InkluExklu} runs over the weights 
$W_s(t)$ 
of all subclusters $s$ 
of $c$. Recall that for the (quasi-)one-dimensional geometry considered 
here, all subclusters are again just chains or ladders of finite size and Eq.\ 
\eqref{Eq::InkluExklu} can be organized rather easily. 

Within the NLCE, the contribution of each cluster is evaluated 
numerically exact. 
Thus, in practice, 
the series in Eq.\ \eqref{Eq::NLCE} has to be truncated to a maximum cluster 
size $c_\text{max}$ which remains computationally feasible. This 
truncation in turn leads to a breakdown of convergence of $C(t)$ at some time 
$\tau$, where a larger $c_\text{max}$ leads to a longer $\tau$, see also 
\cite{Richter2019, White2017}.
Thanks to the combination of NLCE with DQT (cf.\ Sec.\ \ref{Sec::DQT} and 
\cite{Richter2019}), we can evaluate $\langle j(t) j 
\rangle_\text{eq}^{(c)}$ on large clusters beyond the range of ED, and obtain 
$C(t)$ in the thermodynamic limit for rather long times. 

Eventually, let us note that while we here 
focus on
$\beta = 0$, both DQT and NLCE 
allow for accurate calculations of $C(t)$ at $\beta > 0$ as well 
\cite{Steinigeweg2014, Steinigeweg2015, Richter2019}.
\begin{figure}[tb]
 \centering
 \includegraphics[width=1\columnwidth]{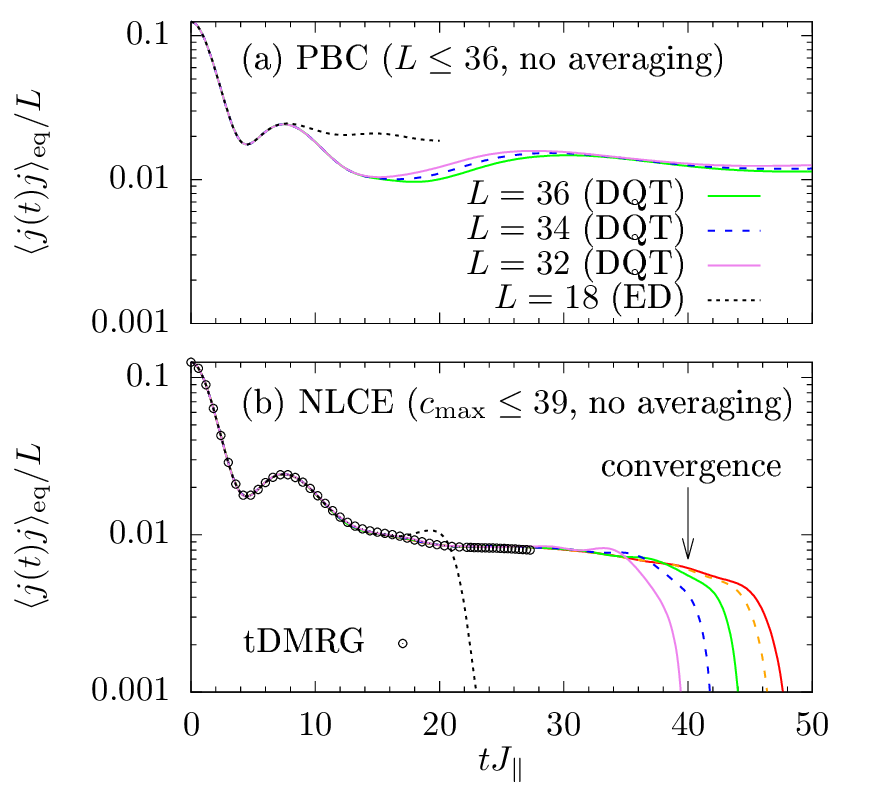}
 \caption{(Color online) (a) $C(t)$ for $J_\perp = 0$ and $\Delta = 1$ at 
$\beta = 0$, obtained by 
ED ($L = 18$) and DQT ($L \leq 39$) for PBC. $L$ here denotes 
the length of a single chain. (b) $C(t)$ obtained by NLCE up to expansion 
order $c_\text{max} \leq 39$. As a comparison, we depict tDMRG data 
\cite{Karrasch2015}.}
 \label{Fig1}
\end{figure}

\subsection{Results for unperturbed dynamics}\label{Sec::Unperturbed}

Now, let us study $C(t)$ in the unperturbed 
system ${\cal H}_0$. For vanishing anisotropy $\Delta = 0$, the spin current 
$j$ 
is exactly conserved in
the unperturbed system, $[{\cal H}_0,j] = 0$. 
As a consequence, we know the unperturbed dynamics for $\Delta = 0$ exactly and 
only need to study the case $\Delta = 1$ numerically, where ${\cal 
H}_0$ corresponds to two separate Heisenberg chains. 
In the remainder of Sec.\ \ref{Sec::Unperturbed}, we change 
the notation and denote by $L$ the length of a single chain (and not of the 
ladder). 
\begin{figure}[tb]
 \centering
 \includegraphics[width = 1\columnwidth]{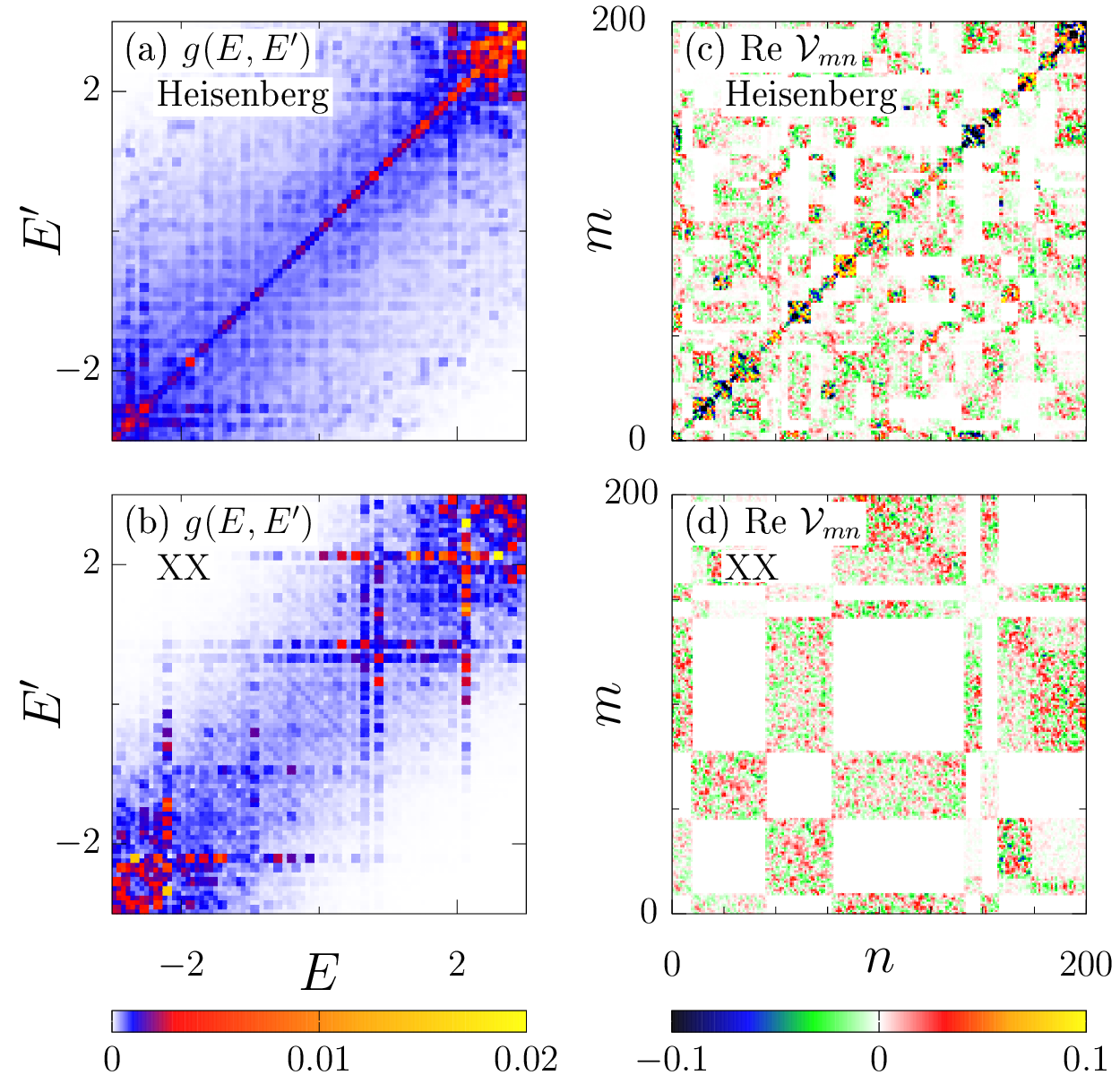}
 \caption{(Color online) Matrix structure of ${\cal V}$ in the eigenbasis of 
${\cal H}_0$ in the symmetry subsector with magnetization $S^z = -1$, momentum 
$k = 2\pi/L$, and even parity $p = 1$, both for the Heisenberg ladder (top) 
and the XX ladder (bottom). The length of the ladder is $L = 9$ in all cases.}
 \label{Fig2}
\end{figure}

In Fig.\ \ref{Fig1}~(a), $\langle j(t) j \rangle_\text{eq}/L$ is shown 
for 
periodic boundary conditions (PBC), obtained by ED ($L = 18$) and DQT ($L = 32, 
34, 36$) \cite{Steinigeweg2014, Richter2018}. 
While the curves for 
different $L$ 
coincide at short times, 
the ED curve starts to deviate from the DQT data for $t \gtrsim 8$. Moreover, 
for $t \gtrsim 20$, $C(t)$ takes on an approximately constant value which 
decreases with increasing $L$ \cite{Steinigeweg2014}.

Next, NLCE results for $C(t)$ are shown in Fig.\ 
\ref{Fig1}~(b) for various expansion orders $c_\text{max} \leq 
39$. For 
increasing 
$c_\text{max}$, we find that $C(t)$ is converged up to longer and longer 
times, until the expansion eventually breaks down. [Note that, for times above 
the convergence time, the data for $C(t)$ obtained by NLCE (i) has no physical 
meaning anymore and (ii) can for instance become negative, which leads to the 
discontinuity of the curves in the semilogarithmic plot used.]
As a comparison, Fig.\ \ref{Fig1}~(b) also shows data obtained by the 
time-dependent density 
matrix renormalization group (tDMRG) \cite{Karrasch2015, Kennes2016}.
Apparently, tDMRG and NLCE 
agree perfectly 
for times $t \lesssim 27$. Moreover, for the largest $c_\text{max} = 39$ 
considered by us, 
the NLCE data is converged up to times $t 
\approx 
40$. This fact demonstrates that the combination of DQT and NLCE provides a 
powerful numerical approach to real-time correlation functions, and 
compared to Fig.\ \ref{Fig1}~(a), also outperforms standard finite-size 
scaling on short to intermediate time scales.

Note that the determination of the unperturbed dynamics extends earlier results 
of Ref.\ \cite{Richter2019} and is an important building block of this paper 
in order to evaluate the prediction from the TCL formalism in Sec.\ 
\ref{Sec::Perturbed}.  

\subsection{Matrix structure of the perturbation}\label{Sec::MatrixStructure}

Before discussing the real-time dynamics of $C(t)$ in the presence of ${\cal 
V}$, let us study the matrix structure of the realistic perturbation ${\cal V}$ 
from Eq.\ 
\eqref{Eq::Perturb} in the eigenbasis of ${\cal H}_0$.
To this end, we restrict 
ourselves to a single symmetry subsector with magnetization $S^z = -1$, 
momentum $k = 2\pi/L$, and even parity $p = 1$ to eliminate trivial symmetries. 
(Both ${\cal H}_0$ and ${\cal 
V}$ are entirely real in this case.)

First, we employ a 
suitable coarse graining according to \cite{Richter2019_3}
\begin{equation}\label{Eq::gEE}
g(E,E') = \frac{\sum_{mn} |{\cal V}_{mn}|^2 D(\bar{E})}{D(E) D(E^\prime)} \, ,
\end{equation}
where the sum runs over matrix elements ${\cal V}_{mn}$ in two energy 
shells of width $2\delta E$, $E_n \in [E - \delta E,E + \delta E]$ and $E_m \in 
[E^\prime -\delta E,E^\prime+\delta E]$. $D(E)$, $D(E')$, and $D(\bar{E})$ 
denote the number of states in these energy windows with mean energy $\bar{E} 
= (E + E^\prime)/2$.  
In Figs.\ \ref{Fig2}~(a) and (b), this coarse-grained structure of 
${\cal V}$ is shown. Both for $\Delta = 1$ 
and $\Delta = 0$, we find that ${\cal V}$ is a banded matrix with 
more spectral weight close to the diagonal.
However, especially in the case of the XX ladder, $g(E,E')$ is not homogeneous 
within this band, but rather exhibits some fine structure. 

For a more detailed 
analysis, a close-up of $200 \times 200$ individual matrix elements ${\cal 
V}_{mn}$ is 
shown in Figs.\ \ref{Fig2}~(c) and (d). We find that there is a 
coexistence between regions where the ${\cal V}_{mn}$ appear to be random, and 
regions where the ${\cal V}_{mn}$ vanish (e.g.\ due to additional conservation 
laws). Moreover, in the case of the XX ladder, these regions are more extended. 

While it becomes obvious from Fig.\ \ref{Fig2} that the 
perturbation ${\cal V}$ is certainly not an entirely random 
matrix in the eigenbasis of ${\cal H}_0$ (which is also 
integrable such that the ETH is generally not expected to apply), we here 
refrain from a more detailed 
analysis of the residual correlations between the matrix elements. 
Nevertheless, given the overall banded structure of ${\cal V}$ and its apparent 
partial randomness, it is reasonable that our derivations from Sec.\ 
\ref{Sec::TCL} can be relevant for this realistic model.   

\subsection{Comparison between perturbed dynamics and exponential 
damping}\label{Sec::Perturbed}

\subsubsection{XX ladder}\label{Sec::XXLad}

\begin{figure}[tb]
 \centering
 \includegraphics[width=1\columnwidth]{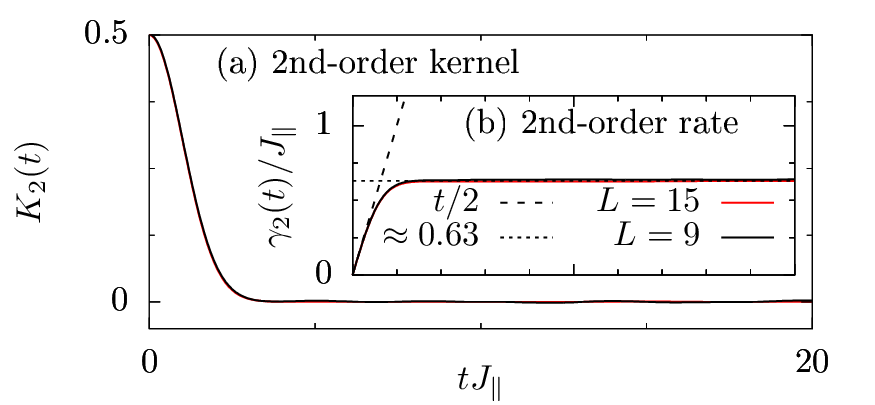}
 \caption{(Color online) (a) and (b) Second-order 
kernel $K_2(t)$ and decay rate $\gamma_2(t)$ for the XX ladder.}
 \label{Fig3}
\end{figure}

Next, we come to the actual discussion of $C(t)$ in spin 
ladders. ($L$ now denotes the length of the ladder.) Given the decomposition 
of ${\cal H}_0$ and 
${\cal V}$ in Eq.\ \eqref{Eq::Ham} and the choice of the projector in Eq.\ 
\eqref{Projecj}, we can directly apply the TCL formalism 
to the decay of $C(t)$ in this model.
First, we consider the case $\Delta = 0$, i.e, the XX ladder.
To start our analysis, we present in Figs.\ 
\ref{Fig3}~(a) and (b) the second-order kernel $K_2(t)$, 
\begin{equation}\label{Eq::K2Exact}
 K_2(t) = \frac{\text{Tr}\lbrace i[j,{\cal V}_I(t)]i[j,{\cal 
V}]\rbrace}{\text{Tr}\lbrace j^2 
\rbrace }\ .   
\end{equation}
Due to $[j,{\cal H}_0] = 0$, or in cases where the dynamics in ${\cal H}_0$ is
slow compared to the dynamics in ${\cal H}$ \cite{Steinigeweg2010}, this kernel 
simplifies to 
\begin{equation}\label{Eq::K2Approx}
  K_2(t) = \frac{\text{Tr}\lbrace i[j,{\cal V}_I](t)i[j,{\cal 
V}]\rbrace}{\text{Tr}\lbrace j^2 
\rbrace }\ .
\end{equation}
The corresponding decay 
rate $\gamma_2(t)$ reads, 
\begin{equation}
 \gamma_2(t) = \int_0^t \text{d}t' 
K_2(t')\ . 
\end{equation}
Comparing data for $L = 9, 15$, we observe that finite-size effects are 
negligible, 
and $\gamma_2(t) \approx 0.63$ becomes essentially constant for 
times $t \gtrsim 2$. Note that, since the XX chain can be brought into a 
quadratic form, $K_2(t)$ and $\gamma_2(t)$ could in principle even be obtained 
analytically for this 
particular case (see also \cite{Steinigeweg2011}).
\begin{figure}[tb]
 \centering
 \includegraphics[width=1\columnwidth]{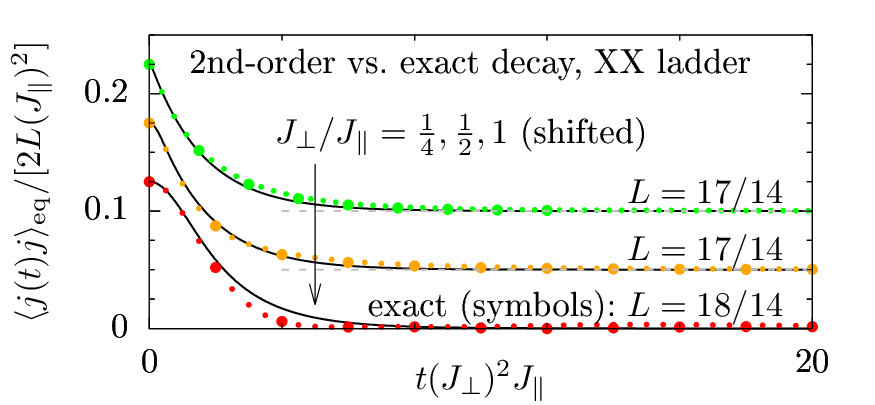}
 \caption{(Color online) $C(t)$ in XX ladders with several interchain 
couplings 
$J_\perp$ 
at $\beta = 0$, obtained by DQT
for two different $L \leq 18$ (small and large  
symbols). The curves indicate the 
second-order TCL prediction \eqref{Eq::Prediction}.}
 \label{Fig4}
\end{figure}

To proceed, Fig.\ \ref{Fig4} presents numerical data for the current 
autocorrelation function $C(t)$ for XX ladders with two different system sizes 
$L \leq 18$ and different coupling ratios 
$J_\perp/J_\parallel = 1/4, 1/2, 1$ (symbols) \cite{Steinigeweg2014_2}, i.e., 
weak and strong values of 
the perturbation. Note that the data is vertically shifted 
for better 
visibility. Moreover, while the data is here obtained by DQT, we present NLCE 
data for the nonintegrable ladder model in Appendix \ref{AppendixA}. 
As a comparison, the curves in Fig.\ \ref{Fig4} indicate our main result 
\eqref{Eq::ExpDec}, i.e., the lowest-order prediction from the TCL formalism 
which reads
\begin{equation}\label{Eq::Prediction}
 C(t) = C_0(t) \exp\left[ -J_\perp^2 \int_0^t \text{d}t' \gamma_2(t') \right]\ 
.
\end{equation}
Recall that $j$ is exactly conserved in the unperturbed system, i.e., $C_0(t) 
= \text{const}.$, and the decay of $C(t)$ is solely due to ${\cal V}$. 
In Eq.\ \eqref{Eq::Prediction}, we take into account the full time 
dependence of the damping rate $\gamma_2(t)$.
Namely, due to the linear growth of $\gamma_2(t)$ at short times, 
Eq.\ \eqref{Eq::Prediction} leads to a Gaussian damping for $t 
\lesssim 1$, 
and turns into a conventional exponential damping for longer $t$, 
\begin{equation}
 C(t) \propto \begin{cases}
               e^{-aJ_\perp^2 
t^2} & t 
\lesssim 1 \\
e^{-bJ_\perp^2 t} & t > 1
              \end{cases}\ .  
\end{equation}

As an important result, we find that 
Eq.\ \eqref{Eq::Prediction} describes the actual decay of  
$C(t)$ remarkably well, albeit the agreement is certainly better 
for smaller $J_\perp/J_\parallel$.
In this context, let us emphasize that for time-dependent problems, 
a truncation to lowest order is in general (for nonrandom matrices) not 
meaningful, even if the perturbation is small. For small perturbations, 
the relevant time scales are long and higher orders can become relevant on 
these 
time scales. 
Thus, the good agreement in Fig.\ \ref{Fig4} confirms that our 
derivations in the context of Eq.\ \eqref{Eq::ExpDec} can be relevant also for 
realistic perturbations and quantum many-body systems.
A more detailed comparison and discussion can be found further below 
in Sec.\ \ref{Sec::FurtherComp}.

\subsubsection{XXX ladder}\label{Sec::XXXLad}

\begin{figure}[tb]
 \centering
 \includegraphics[width=1\columnwidth]{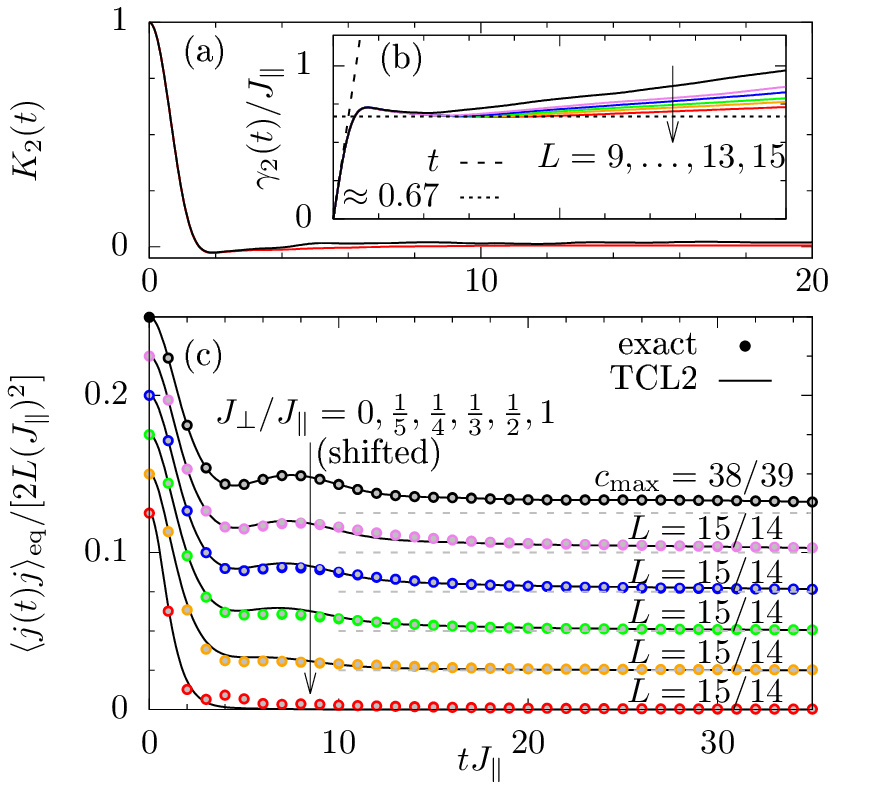}
 \caption{(Color online) (a) and (b) Analogous data as in Fig.\ 
\ref{Fig3}, but now for 
the 
XXX ladder. (c) Analogous data as in Fig.\ \ref{Fig4}, but now for 
the 
XXX ladder.}
 \label{Fig5}
\end{figure}

In order to corroborate our findings further, let us study another but similar 
model. Namely, we consider the dynamics of $C(t)$ for the case of $\Delta = 1$, 
i.e., the XXX ladder.

The second order kernel $K_2(t)$ and the corresponding damping rate $\gamma_2(t)$
are shown in Figs.\ \ref{Fig5}~(a) and (b) for various $L \leq 15$. Here, we again 
use for the kernel $K_2(t)$ the simplified version in Eq.\ \eqref{Eq::K2Approx},
which is strictly valid for $[j,{\cal H}_0]= 0$ only. Despite having $[j,{\cal H}_0]
\neq 0$ in the case of the XXX ladder, it turns out that the finite-size scaling of
this simplified form is much more favorable compared to the exact form in Eq.\ 
\eqref{Eq::K2Exact} (not depicted here) and, as discussed below, allows for an 
accurate description of the decay process.  

Next, in Fig.\ \ref{Fig5}~(c), the autocorrelation 
function $C(t)$ is shown for several values of the 
interchain coupling $J_\perp/J_\parallel = 1/5, \dots ,1$.
We find that data for two different system 
sizes $L = 14,15$ (symbols) nicely coincide with each other, 
i.e., at least on the time scales depicted trivial finite-size effect are 
negligible. 
This can be understood by, e.g., the onset of quantum chaos 
in the nonintegrable ladder and the smaller mean free path of spin 
excitations. (Additional NLCE data for the XXX ladder can be found in 
Appedix \ref{AppendixA}.) 

Analogous to our discussion in the context of Fig.\ \ref{Fig4}, let us now 
compare this temporal decay of $C(t)$ to the prediction of an 
exponential damping. To this end, the unperturbed correlation function 
$C_0(t)$ is exponentially damped 
according to Eq.\ \eqref{Eq::Prediction}. Note that, as an important difference 
to the case $\Delta = 0$, we now have a situation where the unperturbed 
dynamics $C_0(t)$ explicitly depends on time [see Fig.\ \ref{Fig1}~(b)]. 

Similar to the case of the XX ladder, we find that Eq.\ 
\eqref{Eq::Prediction} agrees very well with the exact $C(t)$ for all 
values of $J_\perp$ shown here, even when the perturbation is not weak. Let 
us stress that there is no free parameter involved. 

\begin{figure}[tb]
\centering
\includegraphics[width=1\columnwidth]{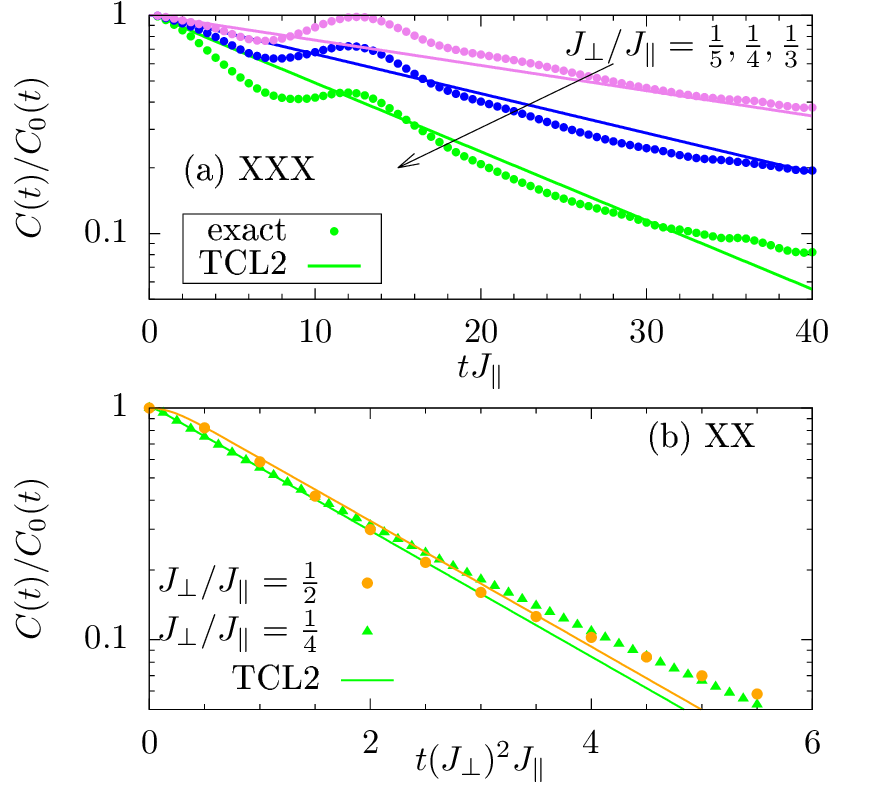}
\caption{(Color online) Ratio $C(t)/C_0(t)$ between the autocorrelation 
function $C(t)$ in ladders with different $J_\perp$ and 
the unperturbed dynamics $C_0(t)$. The 
exact dynamics (symbols) is compared to the prediction from the second-order 
TCL formalism, cf.\ Eq.\ \eqref{Eq::Prediction}. (a) XXX 
ladder; (b) XX 
ladder.}
\label{Fig6}
\end{figure}

\subsubsection{Detailed analysis of agreement and 
discussion}\label{Sec::FurtherComp}

Eventually, for a more detailed analysis, Fig.\ \ref{Fig6} shows the ratio 
$C(t)/C_0(t)$ between the perturbed and the unperturbed dynamics on a 
logarithmic scale, both for the XXX ladder [Fig.\ \ref{Fig6}~(a)] and the XX 
ladder [Fig.\ \ref{Fig6}~(b)]. Again, we compare the exact dynamics obtained 
by DQT and NLCE 
(symbols) to the prediction from the second-order TCL formalism (curves), 
i.e., the curves in Fig.\ \ref{Fig6} now represent the exponential 
damping term $\exp[-J_\perp^2 \int_0^t \gamma_2(t')\text{d}t']$. 

For the XXX ladder, we find that while the exact dynamics exhibits some 
additional oscillations at intermediate times, the overall decay is 
convincingly 
captured by the TCL approach for the interchain couplings $J_\perp=1/5, 1/4, 
1/3$ shown here.
In particular for XX ladders with $J_\perp = 1/4, 1/2$, we find a very good 
agreement 
between the exact 
dynamics and the second-order prediction, at least for times $t(J_\perp)^2 \leq 
4$. 

Note that for time scales and coupling ratios beyond the ones shown in 
Fig.\ \ref{Fig6}, especially in the 
long-time limit where finite-size effects still play a role, the agreement 
between the the 
lowest-order prediction and the exact dynamics becomes less 
convincing. Nevertheless, let us emphasize that for short to intermediate 
time scales (where most of the 
decay happens), an exponential damping is a 
convincing description of the relaxation dynamics of a perturbed quantum 
many-body systems.


\section{Conclusion}\label{Sec::Conclu}

How does the expectation-value dynamics of some operator changes under a
perturbation of the system's Hamiltonian? Based on projection operator 
techniques, we have answered this question for the case of a perturbation with 
random-matrix structure in the eigenbasis of the unperturbed Hamiltonian. 
As a main result, we have unveiled that such a (small) perturbation
yields an exponential damping of the original reference dynamics, consistent 
with recent results in Refs.\ \cite{Dabelow2019, Nation2019}.   

In addition, we have numerically confirmed that our 
findings can in some cases be readily applied to generic quantum 
many-body systems. Specifically, we have studied the decay of current 
autocorrelation functions in spin-$1/2$ ladder systems, where the rungs of the 
ladder are treated as a perturbation to the otherwise uncoupled legs. For this 
example, we have illustrated that even a truncation to 
second order in the perturbation still provides a convincing description of the 
main part of the decay process, also in cases where the perturbation is not 
weak.

While we have shown that for the specific spin-ladder model under 
consideration, the matrix representation of the perturbation ${\cal 
V}$ in the eigenbasis of ${\cal H}_0$ at least partially exhibits random 
segments, it is certainly questionable that realistic physical 
perturbations can be generally described by entirely random matrices.
Therefore, a promising direction of future research includes the identification 
of relevant substructures, as well as a better understanding of the pertinent 
correlations between matrix elements \cite{Foini2019}. 


\subsection*{Acknowledgments} 
This work has been funded by the Deutsche Forschungsgemeinschaft (DFG) - 
Grants No.\ 397107022 (GE 1657/3-1), No.\ 397067869 (STE 
2243/3-1), No.\ 355031190 - within the DFG Research Unit FOR 2692. 
Additionally, we gratefully acknowledge the computing time, granted by the 
``JARA-HPC Vergabegremium'' and provided on the ``JARA-HPC Partition'' part of 
the supercomputer ``JUWELS'' at Forschungszentrum J\"ulich. 


\appendix

\section{NLCE data for another integrable model}
\begin{figure}[tb]
\centering
\includegraphics[width=1\columnwidth]{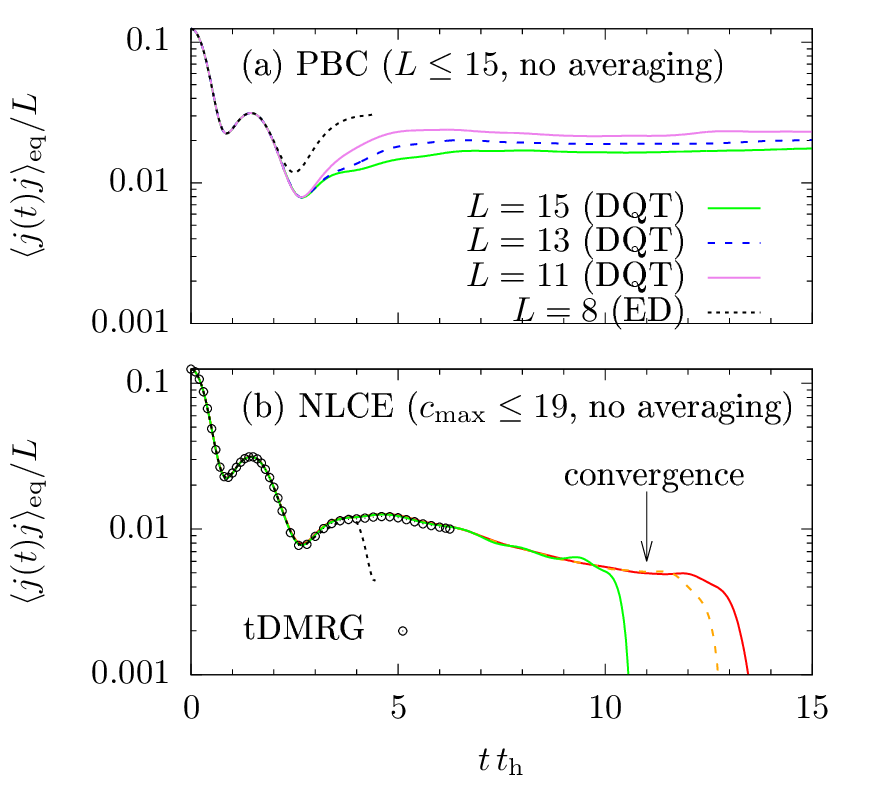}
\caption{(Color online) (a) Current autocorrelation $C(t)$ in the 
Fermi-Hubbard chain ($U = 4$), obtained by ED ($L = 8$) and DQT ($L 
\leq 19$) at $\beta = 0$ for  periodic boundary conditions (PBC). (b) $C(t)$ 
obtained by NLCE up to expansion order $c_\text{max} \leq 19$. As a comparison, 
we depict data from the time-dependent density matrix renormalization group 
(tDMRG) 
\cite{Karrasch2014}.
}
\label{Fig7}
\end{figure}

In the main text, we have used a combination of DQT and NLCE to calculate the 
unperturbed dynamics of the spin-current autocorrelation function in the 
integrable 
spin-$1/2$ Heisenberg chain. This combination has allowed us to obtain 
numerically exact information on rather long time scales, which cannot be 
reached in direct calculations in systems with periodic or open boundary 
conditions, due to significant finite-size effects. To illustrate that this 
combination of DQT and NLCE can yield also for other integrable models the 
reference dynamics with a similar quality, we show additional data for the 
Fermi-Hubbard chain, described by the Hamiltonian ${\cal H} = \sum_{l=1}^L h_l$,
\begin{equation}
h_l = -t_\text{h} \sum_{s=\downarrow, \uparrow} (a_{l,s}^\dagger a_{l+1,s} +
\text{H.c.}) + U (n_{l,\downarrow} - \frac{1}{2})(n_{l,\uparrow} - \frac{1}{2})
\, ,
\end{equation}
where the operator $a_{l,s}^\dagger$ ($a_{l,s}$) creates (annihilates) at site
$l$ a fermion with spin $s$, $t_\text{h}$ is the hopping matrix element, and $L$
is the number of sites. The operator $n_{l,s}$ is the local occupation number 
and $U$ is the strength of the on-site interaction. For this model, we consider 
the particle current $j = \sum_{l=1}^L j_l$,
\begin{equation}
j_l = - t_\text{h} \sum_{s=\downarrow, \uparrow} (i a_{l,s}^\dagger a_{l+1,
s} + \text{H.c.}) \, ,
\end{equation}
and summarize our numerical results for the corresponding autocorrelation 
function with
$U = 4$ in Fig.\ \ref{Fig7}. Apparently, the situation is like the one in 
Fig.\ \ref{Fig1} of the main text. On the one hand, in direct calculations 
with periodic boundary conditions, strong finite-size effects set in at 
short times, even for quite large $L$. On the other hand, NLCE for the largest 
expansion order $c_\text{max}$ is converged to substantially longer times. Even 
though not shown explicitly, we have checked that a good convergence is also 
reached for $U = 8$. We thus expect that a perturbative analysis, as presented 
in this work, can be carried out for a wide class of quantum many-body systems, 
which we plan to do in detail in future work.
\begin{figure}[b]
\centering
\includegraphics[width=1\columnwidth]{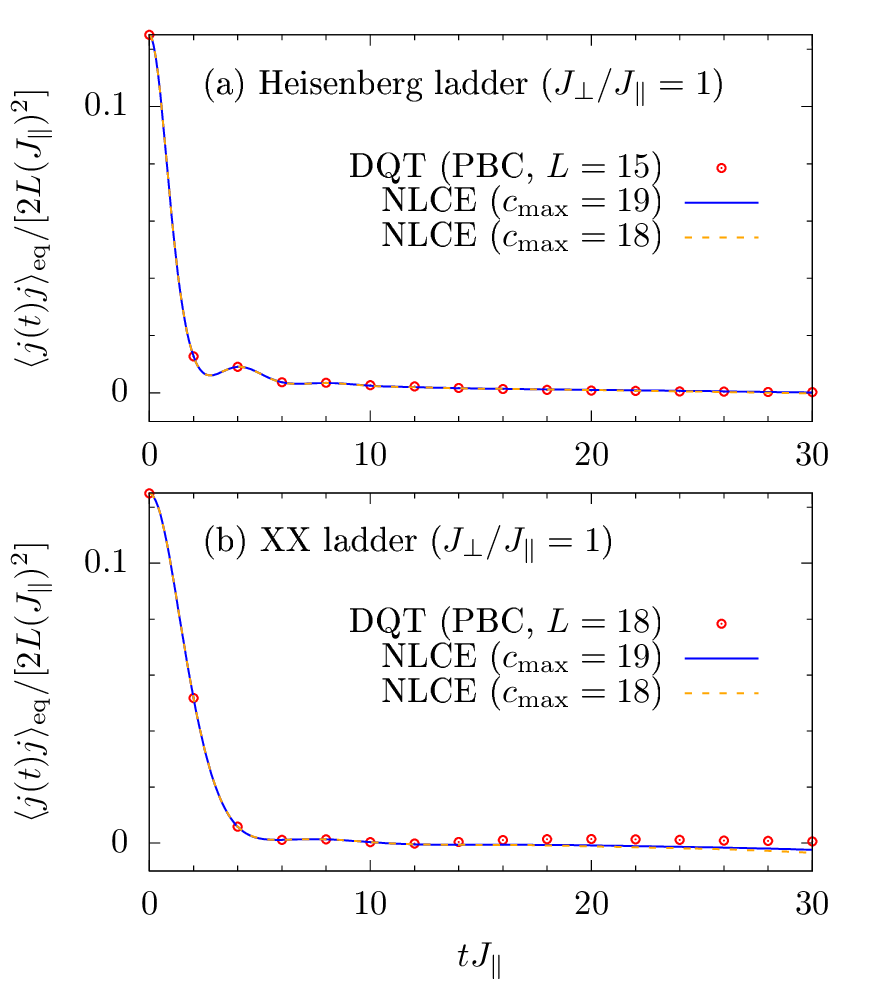}
\caption{(Color online) Current autocorrelation $C(t)$ for the (a) Heisenberg 
ladder and (b) XX ladder, in both cases for $J_\perp/J_\parallel = 1$. DQT data 
for periodic boundary conditions, as shown in the main text, is compared to 
NLCE data for two expansion orders $c_\text{max} = 18$ and $19$.
}
\label{Fig8}
\end{figure}

\section{NLCE data for nonintegrable models}\label{AppendixA}

While it is certainly possible to use NLCE also for nonintegrable models, 
finite-size effects in direct calculations are much less pronounced in these 
models, as evident from Figs.\ \ref{Fig4} and \ref{Fig5} of the main text. This 
is why we have not shown corresponding NLCE data in these figures and instead 
relied on pure DQT data for systems with periodic boundary conditions. To 
demonstrate that these DQT data are indeed in excellent agreement with NLCE 
data, we show in Fig.\ \ref{Fig8} corresponding numerical results for the 
XXX and XX ladder, where we have chosen $J_\perp/J_\parallel 
= 1$ in both cases.


\end{document}